\documentclass[aps,prl,twocolumn,groupedaddress,showpacs,showkeys]{revtex4-1}
\usepackage[dvips]{graphicx,color}

\begin{document}



\title{Volatility-dependent damping of evaporation-driven B\'enard-Marangoni instability}


\author{Fabien Chauvet}
\email[]{fchauvet@gmail.com}
\author{Sam Dehaeck}
\author{Pierre Colinet}
\email[]{pcolinet@ulb.ac.be}
\homepage[]{http://www.tips-ulb.be/}
\affiliation{Universit\'e Libre de Bruxelles, TIPs (Transfers, Interfaces and Processes), C.P. 165/67, 50 Av. F.D. Roosevelt, 1050 Brussels, Belgium}


\date{\today}

\begin{abstract}
The interface between a pure liquid and its vapor is usually close to saturation temperature, hence strongly hindering any thermocapillary flow. In contrast, when the gas phase contains an inert gas such as air, surface-tension-driven convection is easily observed. We here reconcile these two facts by studying the corresponding crossover experimentally, as a function of a new dimensionless number quantifying the degree of damping of interfacial temperature fluctuations. Critical conditions are in convincing agreement with a simple nonlocal one-sided model, in quite a range of evaporation rates.
\end{abstract}

\pacs{47.20.Dr, 47.54.-r, 68.03.Fg}
\keywords{liquid films, evaporation, B\'enard-Marangoni instability, pattern formation, wavelength selection}

\maketitle
When a layer of volatile liquid is exposed to a dry non-condensable (or inert) gas, evaporation occurs and the heat needed
for the phase change induces a cooling of the liquid/gas interface. At large enough thickness and/or evaporation
rate (i.e. large enough Marangoni number $Ma$), surface-tension-driven convection is observed \cite{Mancini}, 
often in the form of hexagonal patterns with their typical defects \cite{Colinet_book_01}. 
However, the analogy with the classical B\'enard-Marangoni (BM) convection is far from straightforward, 
given a number of complicating issues partly discussed hereafter (see also \cite{Haut_Colinet_05}).
Among these, saturation of the gas phase should play an essential role, given that surface temperature 
variations are a priori excluded in a one-component liquid-vapor system (indeed, except for extremely small liquid
depths, the interface should be close to local equilibrium \cite{Colinet_book_01}). Another extreme is the non-volatile 
case, where experiments \cite{Schatz} have confirmed Pearson's theoretical value of $Ma_c \simeq 80$ for the 
critical Marangoni number \cite{Pearson}. Hence, $Ma_c$ should somehow depend on the liquid volatility, a convenient
measure of which is the mole fraction $\omega_\Sigma=p_{\rm sat}/p_{g}$, i.e. the ratio of the saturation pressure $p_{\rm sat}$ at 
ambient temperature to the total gas pressure $p_{g}$. Studying the crossover between the nonvolatile regime ($\omega_\Sigma \ll 1$)
and the pure vapor case ($\omega_\Sigma \rightarrow 1$) is the main goal of the present Letter.
 
In addition to its fundamental interest, pattern formation in drying liquid layers or thin films presently
lacks quantitative understanding, which is however crucial for many techniques such as polymer coating \cite{Bassou}
or nanoparticle deposition \cite{Maillard}. An essential step towards the analysis of the wide variety 
of obtained patterns is a correct evaluation of the supercriticality $\epsilon=(Ma-Ma_c)/Ma_c$, which clearly
requires an accurate determination of $Ma_c$. Another challenging objective is to understand the rich nonlinear
dynamics of evaporation-driven BM patterns in order to control and even optimize deposition structures, which 
requires developing simple models of highly supercritical convection. This is the second aim of this 
Letter, which presently focuses on the case of a pure liquid, where convection is driven by thermal effects only. 

In order to understand the mechanism responsible for the damping of interfacial temperature fluctuations, 
as a function of the volatility, let us consider a flat interface (at $z=0$, with $z$ pointing to the gas) at 
temperature $T_\Sigma$, where evaporation occurs at a mass flux density $J$ (in kg/m$^2$s). Hereafter,
a subscript $\Sigma$ will denote a quantity evaluated at the interface, the gas mixture is taken to 
be perfect, its total pressure $p_g$ is supposed to be constant and uniform (small dynamic viscosity), 
and the inert gas (say, air) is not absorbed into the liquid. This implies (see e.g. \cite{Haut_Colinet_05})
\begin{equation}
	J=-\frac{D M_v}{R\, T_{\Sigma}}\left.\frac{\partial_z p_v}{1-\omega}\right|_{z=0},
	\label{eq_J}
\end{equation}
where $D$ is the vapor-air diffusion coefficient, $M_v$ is the molar mass of the vapor, $p_v$ is the partial pressure of vapor in the gas phase, $\omega=p_v/p_g$ its mole fraction, and $R$ is the ideal gas constant. In addition, the energy balance at the interface reads
\begin{equation}
	\left.\frac{\partial T_l}{\partial z}\right|_{z=0}=-\frac{J {\cal L}}{\lambda_l}+\frac{\lambda_g}{\lambda_l}\left.\frac{\partial T_g}{\partial z}\right|_{z=0},
	\label{eq_thermalflux}
\end{equation}
where $T_l$ and $T_g$ are respectively the liquid and gas temperatures, ${\cal L}$ is the latent heat of vaporization, while $\lambda_l$ and $\lambda_g$ are respectively the liquid and gas thermal conductivities (with $\lambda_g \ll \lambda_l$ in general).

We now consider fluctuations (denoted by tilded quantities) around a particular steady (or quasi-steady) state
distinguished by a superscript $0$. The determination of this particular state needs not be detailed for the
moment, and in principle, the following reasoning applies to both evaporation ($J^0>0$) and condensation ($J^0<0$).
The interface temperature is written $T_\Sigma=T_\Sigma^0+\widetilde{T}_\Sigma$, and assuming local chemical 
equilibrium at the interface, the corresponding fluctuation of vapor partial pressure there 
reads $\widetilde{p}_{v\Sigma}=p_{\rm sat}' (T_\Sigma^0)\, \widetilde{T}_\Sigma$, where $p_{\rm sat}(T)$ is the 
coexistence (i.e. Clausius-Clapeyron) curve and a prime denotes its derivative. As fluctuations 
satisfy $\nabla^2 \widetilde{p}_v=0$ in the limit of a small P\'eclet number (defined on a typical length 
scale of the fluctuations, assumed to be much smaller than the typical size $H$ of the gas phase) 
and in the quasi-static hypothesis, we 
have $\widetilde{p}_{v {\bf q}}=p_{\rm sat}' (T_\Sigma^0)\, \widetilde{T}_{\Sigma {\bf q}} \,{\rm e}^{-\vert {\bf q} \vert z}$,
where a subscript ${\bf q}$ indicates the Fourier component with wavevector ${\bf q}$ (in the horizontal plane). Similarly, 
one also has $\nabla^2 \widetilde{T}_g=0$, because the Lewis number $Le=D/\kappa_g$ is $O(1)$ 
in the gas ($\kappa_g$ is the gas thermal diffusivity). 
Hence, assuming $T_g=T_l\,(=T_\Sigma)$ at $z=0$, we have $\widetilde{T}_{g {\bf q}}=\widetilde{T}_{\Sigma {\bf q}} \,{\rm e}^{-\vert {\bf q} \vert z}$.

Finally, linearizing Eq. (\ref{eq_J}), we can calculate (the Fourier transform of) the phase change rate fluctuation  
\begin{equation}
 \widetilde{J}_{\bf q}=\vert {\bf q} \vert \frac{D M_v}{R\, T_\Sigma^0} \frac{p_{\rm sat}' (T_\Sigma^0)}{1-\omega_\Sigma^0} \widetilde{T}_{\Sigma {\bf q}}
 \label{eq_Jq}
\end{equation}
where fluctuations of the denominator have been neglected (this is rigorously valid for $\vert {\bf q} \vert^{-1}\ll H$,
as will be shown elsewhere). Then, Fourier transforming the interfacial energy balance (\ref{eq_thermalflux}) 
and grouping terms, we get
\begin{equation}
 \left.\frac{\partial \widetilde{T}_{l {\bf q}}}{\partial z} \right \vert_{z=0} + \alpha \vert {\bf q} \vert \,\widetilde{T}_{\Sigma {\bf q}} =0
 \label{Biot_cond}
\end{equation}
where 
\begin{equation}
 \alpha = \frac{\lambda_g}{\lambda_l} + \frac{{\cal L} D M_v}{\lambda_l R T_\Sigma^0} \frac{p_{\rm sat}' (T_\Sigma^0)} {1-\omega_\Sigma^0}
 \label{alpha}
\end{equation}
Equation (\ref{Biot_cond}) has the form of a Newton's cooling law for liquid temperature fluctuations, with a heat transfer coefficient
depending upon their wavenumber $\vert {\bf q} \vert$ (hence, the physical space expression of Eq. (\ref{Biot_cond}) involves a nonlocal convolution term). The positive dimensionless number $\alpha$ turns out to be an effective gas-to-liquid ratio of thermal conductivities, accounting for the effect of phase change through its second term. In particular, the latter contribution is seen to diverge for $\omega_\Sigma^0 \rightarrow 1$, i.e. in the limit of a pure vapor phase, for which Eq. (\ref{Biot_cond}) implies $\widetilde{T}_{\Sigma {\bf q}} = 0$, i.e. the interface temperature does not fluctuate and remains equal to the saturation (boiling) temperature.

Now, applying Eqs (\ref{Biot_cond}) and (\ref{alpha}) to the modeling of evaporation-driven BM convection in a liquid 
layer of height $e$ much thinner 
than the gas phase thickness $H$, it turns out that Pearson's theory \cite{Pearson} can be straightforwardly applied, using 
an effective Biot number $Bi=\alpha\, k$ where $k=\vert {\bf q} \vert\,e$ is the dimensionless wavenumber. The neutral 
stability threshold is then given by
\begin{equation}
	Ma_k={\frac{16 k (k \cosh [k ] + \alpha k \sinh [k ]) (2 k - \sinh [2 k ])}{4 k^{3} \cosh [k
] + 3 \sinh [k ] - \sinh [3 k ]}}
	\label{eq_Mak}
\end{equation}
The critical Marangoni number $Ma_c(\alpha)$ and the critical wavenumber $k_c(\alpha)$ are then
found by minimizing $Ma_k$ with respect to $k$, and will now be compared to experiments. Note 
finally that the theory just described appears as a particular case of a more general formulation 
(not limited to $e \ll H$) described in \cite{Haut_Colinet_05}, and also based on a one-sided reduction
of the evaporation-driven BM problem by adiabatic slaving of gas phase fluctuations.

In order to test these predictions, accurate experiments were performed by evaporating thin liquid layers 
into ambient air ($T \approx$ 24$^\circ$C) at rest, until the liquid completely disappears. As explained
hereafter, we mostly focus on the moment at which convective patterns disappear in favor of a uniform evaporative
state. Volatile liquids used are Hydrofluoroethers, HFE-7000, 7100, 7200 and 7300 from the company 3M, which have
similar physical properties apart for their saturation pressure $p_{\rm sat}$ (factor of about 2 between two successive HFEs).
HFE-7000 is the most volatile with $p_{\rm sat}(24^\circ$C$)=$ 0.61 bar and HFE-7300 is the less volatile 
with $p_{\rm sat}(24^\circ $C$)=$ 0.06 bar. Other thermodynamic and transport properties used hereafter are found 
in 3M data sheets (available on 3M web site). 

Each experimental run is started by pouring a certain amount of HFE in a cylindrical container to form an approximately 1 mm 
thick liquid layer. The container is made of a PVC cylinder glued by silicone to a 10 mm thick aluminum plate. The height 
of the cylinder is 1 cm, its diameter is 63.5 mm and its thickness is 6 mm. In addition to the effect of volatility (dependent 
on the HFE used), we also vary the evaporation rate independently by changing the ``transfer distance" $H$ in the gas. This is 
accomplished by topping another PVC cylinder (of the same diameter) on the one glued to the plate, wrapping them with a scotch 
tape in order to avoid any vapor leak. Using additional cylinders of various heights allows to set $H$ to 1 cm, 2 cm, 3 cm, 
4 cm and 5 cm.

In these conditions, the evaporation process is limited by diffusion of vapor into air and the evaporation rate $E=J^0 S$ (where $S$
is the container cross section) remains quasi-constant until the layer is too thin and dewetting begins. The liquid film thickness, $e$, is measured by weighting and is deduced from the measured total mass, $m_{tot}$, taking into account the mass of the liquid meniscus against the internal cylinder wall, $m_{men}$, and the mass of the vapor contained in the gas phase above the liquid, $m_v$, such that $e=(m_{tot}-m_{men}-m_v)/\rho_l S$, where $\rho_l$ is the liquid density. Both these contributions cannot be neglected 
because the critical layer thicknesses are generally small. More precisely, $m_{men}$ is theoretically estimated assuming that 
the meniscus is in its hydrostatic equilibrium state and that the liquid is perfectly wetting. In turn, $m_v$ has been 
measured experimentally for each $H$ and each HFE (molecular weight between $200$ and $350$ g/mol) using a suspended thin circular dish filled with liquid and placed very close to the container bottom but without touching the container wall, hence ``simulating'' the presence of a liquid layer. In the worse case (most volatile liquid HFE-7000 and highest container $H=5$ cm), we find $m_v \simeq 0.2\, m_{tot}$ and in the best one (HFE-7300 and $H=1$ cm), we get $m_v \simeq 0.005\, m_{tot}$. The relative uncertainty on the liquid layer thickness measurement is estimated to be lower than 2.5\%. The evaporation rate, $E$, is simply computed from the time derivative of the total mass ($E=-d m_{tot}/dt$) using a linear fit.

As convection in the pure liquid is necessarily associated with temperature variations, we use a Focal Plane Array IR camera-type (Thermosensorik, InSb 640 SM) facing the liquid/gas interface, to follow the time evolution of the whole cellular pattern. IR images and $m_{tot}(t)$ are recorded at a frequency of 1 Hz during the drying of the liquid layer. Typically, the observed sequence is similar to the one obtained in \cite{Mancini}, i.e. convection appears right after filling and the pattern is strongly time-dependent, evolving into more stable hexagonal-like arrangements when $e$ decreases, until the convective state turns into a ``conductive" one. The convection cells do not disappear altogether, however. At a certain moment, a straight front separating convective and conductive states starts to propagate along a horizontal direction (at a constant velocity) until convection completely disappears. Performing specific experiments in which the container tilt angle was intentionally slightly varied showed that this front is merely due to a non-absolute horizontality of the layer (the front velocity decreases when increasing the tilt angle and vice versa). We have chosen to define the critical liquid layer thickness, $e_c$, by $e_c=(e_1+e_2)/2$ where $e_1$ is the measured thickness when the front starts to propagate and $e_2$ the measured thickness when convection has totally disappeared. For the small tilt angles tested, $e_c$ is found to be independent of these small deviations w.r.t. absolute horizontality. 

From the measurement of $e_c$ and $E$, we then estimate the critical temperature difference across the liquid layer, $\Delta T_c$, using the thermal balance (\ref{eq_thermalflux}) with a linear temperature profile in the liquid and neglecting heat coming from the gas phase, such that $\Delta T_c = E {\cal L} e_c/\lambda_l S$. Then, the critical Marangoni number $Ma_c$ is calculated as $Ma_c=-\gamma_T \Delta T_c e_c/\eta_l \kappa_l$, where $\gamma_T$ is the surface tension variation with temperature, $\eta_l$ is the liquid dynamic viscosity and $\kappa_l$ is its thermal diffusivity. The value of $\gamma_T$ has been measured for each HFE by the pendant drop method using the tensiometer Kr\"uss DSA100 with its thermostated chamber, and a thermocouple placed at the syringe tip end to measure the drop temperature accurately. Surface tension has been measured in the range 15-30$^\circ$C, taking special care in order to maintain the drop in a saturated environment (procedure validated by measuring $\gamma_T$ of ethanol).

\begin{figure}[ht]
\centering
		\includegraphics[width=0.45\textwidth]{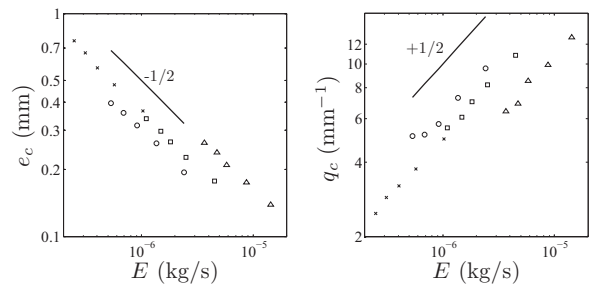}
	\caption{Measured critical liquid thickness $e_c$ (left) and critical wavenumber $q_c$ (right) as a function of the evaporation rate $E$, for different liquids (triangles : HFE-7000, squares : HFE-7100, circles : HFE-7200, crosses : HFE-7300). Straight lines indicate the theoretical scaling laws (see text).}
	\label{fig:scalings}
\end{figure}

According to our model, the value found for $Ma_c$ should only depend on the liquid used (through the value of $\alpha$ 
characterizing the damping of thermal fluctuations at the interface) and not on its evaporation rate $E$. More precisely, 
as $\Delta T_c \sim E \,e_c$ for a given liquid, $Ma_c \sim E\, e_c^2$ should be a constant, leading to the 
scaling $e_c \sim E^{-1/2}$. Apart for some small variations studied hereafter, the size of convection cells
at threshold should be roughly proportional to the depth $e_c$. Hence, the critical 
wavenumber $q_c \sim e_c^{-1} \sim E^{1/2}$. Both these scalings indeed match experimental measurements, 
as shown in Fig. \ref{fig:scalings}. 

\begin{figure}[ht]
\centering
		\includegraphics[width=0.35\textwidth]{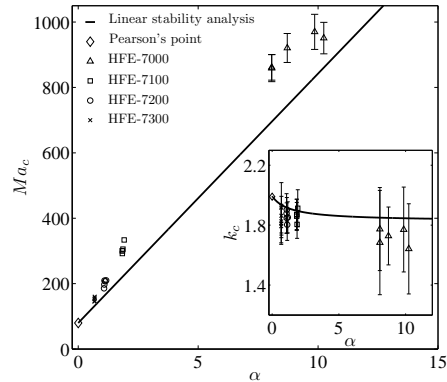}
	\caption{Measured critical Marangoni numbers (symbols) as a function of $\alpha$, for the four different liquids and the five container heights. Inset: corresponding measured critical wavenumbers. Theoretical laws are shown as plain curves.}
	\label{fig:Mac_kc(alpha)}
\end{figure}

Now, in order to fully validate our Pearson-like theory, $\alpha$ is directly computed from Eq. (\ref{alpha}) using our own measured values of $D$ for each HFE, obtained by the Stefan's tube method \cite{Abraham}. The obtained 
values of $Ma_c$ for all the HFEs and for all the container heights $H$ are plotted as a function of $\alpha$ in Fig. \ref{fig:Mac_kc(alpha)}, and compared to the theoretical law $Ma_c(\alpha)$, independent of $E$ as already mentioned.

\begin{figure}[ht]
\centering
		\includegraphics[width=0.3\textwidth]{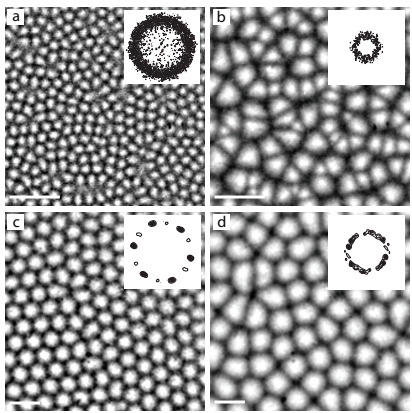}
	\caption{Typical evaporation-driven BM patterns ($H = 3$ cm): 
	(a) HFE-7000, $\epsilon = 0.2$; (b) HFE-7000, $\epsilon = 2$; 
	(c) HFE-7300, $\epsilon = 0.2$; (d) HFE-7300, $\epsilon = 2$.
	White bars are 4 mm long. Top-right insets : contourlines of
	power (Fourier) spectrum.}
	\label{fig:Patterns}
\end{figure}

The inset of Fig. \ref{fig:Mac_kc(alpha)} also shows the dimensionless critical wavenumber of the pattern, $k_c=q_c e_c$, clearly 
independent of $E$ as well. Note that the critical wavenumber $q_c$ (as also plotted in Fig. \ref{fig:scalings}) and the wavenumber $q$ 
(during the convection regime) are measured as the mean position of the fundamental peak in azimuthally averaged FFT spectra of 
IR images (see Fig. \ref{fig:Patterns}). 
Error bars 
on $k_c$ in Fig. \ref{fig:Mac_kc(alpha)} correspond to the width at middle height of the fundamental peak, hence 
indicative of the level of disorder in the pattern (higher at high volatility).

Figure \ref{fig:Mac_kc(alpha)} demonstrates the strong stabilizing effect of the liquid volatility, as well as a quite satisfactory agreement with our simple one-sided theory (given typical uncertainties remaining on some fluid properties and the absence of fitting parameters). 
Note that this actually validates a number of assumptions made in such type of models (see also \cite{Haut_Colinet_05}), 
such as small gas viscous stresses, low P\'eclet numbers in both phases, quasi-steadiness of the approach despite 
the continuously decreasing liquid depth, undeformable interface, absence of temperature discontinuity, ... 
In addition, we emphasize that the simplest form of the  
theory presented here relies on the additional assumption of a large gas thickness $H$ compared to the liquid 
depth $e$. As the length scale of convective fluctuations typically scales with $e$, their penetration depth 
in the gas is of the same order, which in fact allows neglecting the effects of gas density variations and 
diffusion-induced convection (even though both these effects do affect the homogeneous evaporation state, 
hence $J^0$, when $\omega_\Sigma^0$ increases). This will be detailed elsewhere.

\begin{figure}[ht]
\centering
		\includegraphics[width=0.32\textwidth]{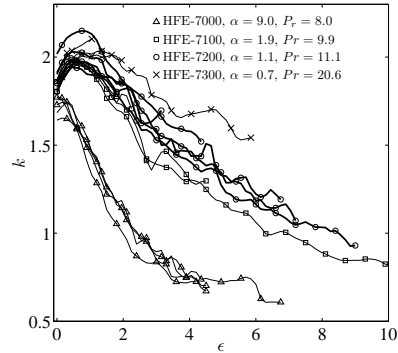}
	\caption{Measured dimensionless wavenumber $k$ versus supercriticality $\epsilon=(Ma-Ma_c)/Ma_c$. For each 
	liquid, evolutions corresponding to all five container heights are represented. $Pr$ is the Prandtl number of the liquids.}
	\label{fig:k}
\end{figure}

To conclude, let us briefly explore nonlinear regimes of evaporative BM convection. Figure \ref{fig:Patterns}
shows that cellular patterns become more regular when the supercriticality $\epsilon=(Ma-Ma_c)/Ma_c$ decreases
(i.e. when time goes on), and that at the same value of $\epsilon$, patterns are more disordered for more volatile 
liquids. This is also confirmed by the corresponding Fourier spectra, from which averaged wavenumbers $k=q\, e$ depicted in 
Fig. \ref{fig:k} are extracted. We first note that the shape of $k(\epsilon)$ curves (including the 
non-monotonic behavior at low $\alpha$) is in nice qualitative agreement with direct simulations 
of \cite{Bestehorn_03}, which however rely on a constant Biot number instead of Eq. (\ref{Biot_cond}).

Interestingly, Fig. \ref{fig:k} also shows that the measured wavenumber evolutions  
are rather independent of the container height (hence on the evaporation rate), 
while they do depend on the liquid used. This clearly has to do with the fact that 
the timescale for liquid depth 
variation, $\tau_e \sim e/\vert \dot{e} \vert$, is always much larger than the thermal time 
scale $\tau_{th} \sim e^2/\kappa_l$ (quasi-steady evolution). However,
the fact that $\tau_e$ turns out to be much smaller than the lateral diffusion 
time $\tau_L \sim L^2/\kappa_l$ ($L$ is the container size) points to a rather 
fast mechanism of wavelength selection, which might be due, at least far from threshold
where the pattern is large-scale ($k\ll k_c$), to the anomalous 
dissipation mechanism described by Eq. (\ref{Biot_cond}). This remains to be studied
however, along with the quite unexplored scenarii of transition to ``interfacial
turbulence" (see also \cite{Bestehorn_03}), for which the one-sided model we propose
should be appropriate in a quantitative sense. Note finally that although validated
here using a B\'enard set-up, the phase-change-induced homogenization mechanism described 
by Eq. (\ref{Biot_cond}) is expected to be generic for other geometries as well (e.g. drops, 
bubbles, ...), at least for sufficiently short-scale interfacial temperature fluctuations.

\begin{acknowledgments}
The authors are grateful to A.A. Nepomnyashchy, A. Oron, M. Bestehorn and R. Narayanan for interesting discussions, and acknowledge financial support of ESA \& BELSPO, EU-MULTIFLOW, ULB, and FRS--FNRS.
\end{acknowledgments}
\bibliography{Chauvet_bib}

\end{document}